\documentclass[epj]{svjour}
\usepackage{graphics}
\usepackage{amsmath}
\usepackage{mathrsfs}
\usepackage{amsfonts}
\usepackage{overpic}
\usepackage{hyperref}
\usepackage{rotating}
\usepackage{cite}
\newcommand{\ket}[1]{|#1\rangle}
\newcommand{\bra}[1]{\langle #1|}
\newcommand{\inp}[2]{\langle #1 | #2\rangle}

\begin{document}
\title{Geometric Entanglement in Valance-Bond-Solid state}
\author{H. T. Cui, C. M. Wang and S. Z. Yuan}
\institute{School of Physics and Electronic Engineering, Anyang
Normal University, Anyang 455000, China \email{cuiht@aynu.edu.cn}}
\date{Received:   / Revised version: \today}

\abstract{Multipartite entanglement, measured by the geometric
entanglement(GE), is discussed for integer spin Valance-Bond-Solid
(VBS) state respectively with periodic boundary condition(PBC) and
open boundary condition(OBC) in this paper. The optimization in the
definition of geometric entanglement can be reduced greatly by
exploring the symmetry of VBS state, and then the fully separable
state can be determined explicitly. Numerical evaluation for GE by
the random simulation is also implemented in order to demonstrate
the validity of the reductions. Our calculations show that GE is
saturated by a finite value with the increment of particle number,
that means that the total entanglement for VBS state would be
divergent under the thermodynamic limit. Moreover it is found that
the scaling behavior of GE with spin number $s$ is fitted as
$\alpha\log (s+\tfrac{\beta}{s}+\gamma)+\delta$, in which the values
of the parameters $\alpha, \beta, \gamma, \sigma$ are only dependent
on the parity of spin $s$. A comparison with entanglement entropy of
VBS state is also made, in order to demonstrate the essential
differences between multipartite and bipartite entanglement in this
model. }

 \maketitle

\onecolumn
\section{Introduction}
The crossover between quantum information and statistical mechanics
fertilizes a distinct insight into the many-body effects and
intrigues extensive interests\cite{afov07, osterloh08}. Especially
inspired by the pioneer works by Osterloh, {\it et.al.}\cite{oaff02}
and Osborne and Nielsen\cite{on02}, quantum entanglement in
many-body systems has received great attention, especially on the
critical systems\cite{afov07, ecp08}. This interest can be
attributed to the following points. It is well known that the
general properties of critical systems can be obtained by
identifying the scaling behavior of the singularity for correlation
functions, e.g. the correlation lengths and related critical
exponents. Furthermore the vanishing of energy gap between the
ground state and the first exciting state may lead to the long-range
correlation even for infinitely separated particles at zero
temperature. Thus it is conjectured that quantum entanglement, a
special quantum correlation, could also display likewise behaviors
at critical points. Some important progresses have been made in this
direction. The recent studies for the block entanglement entropy
have displayed the logarithmical divergency with the block size at
critical points, and the universal area law has also been
constructed exactly in one-dimensional spin-chain systems (see
Ref.\cite{ecp08} and references therein). Furthermore the
multipartite entanglement in many-body systems has also been shown
to be maximal or to display sharp changes closed to the critical
points\cite{multi, odv08, wei05}.

From the point of quantum information, the long-range correlation
implies that a stable quantum channel could be constructed easily
for the transfer of quantum information\cite{tele}. For
half-odd-integer antiferromagnetic Heisenberg spin systems,
Lieb-Schultz-Mattis(LSM) theorem\cite{lsm61} states that there is no
energy gap between the ground state and the first excited state
under thermodynamical limit, and the long-range order exists for any
finite amount of anisotropy couplings. Then the long-distance
quantum teleportation can be implemented in this system\cite{tele,
gi09}. However the situation is different for integer spin-chain
systems. Haldane's conjecture claims that there would be a finite
excitation gap above the ground state for integer-spin
antiferromagnetic Heisenberg chains, and then two-point correlation
decays exponentially with the distance between two
particles\cite{haldane}. In order to verify this interesting issue,
Affleck-Kennedy--Lieb-Tasaki(AKLT) model was constructed, whose
ground state is the so called valence-bond-solid(VBS)
state\cite{aklt}. And the two-point correlation functions for VBS
state has been shown to decay exponentially  because of the advent
of the energy gap above the ground state \cite{arovas88}.
Consequently it is conjectured that quantum entanglement would be
absent between two long distance separated particles in AKLT model.
But the fact beats all. Verstraete, {\it et.al.,} demonstrate that
the maximal entanglement between two infinitely separated particles
can be founded even in the AKLT model by only imposing local
operations on the rest of the particles\cite{verstraete04}. This
finding enforces the reconsideration of the connection between
quantum entanglement and correlation in many-body
systems\cite{venuti06, venuti05}.

Although the absence of long-range correlation in AKLT model, den
Nijs and Rommelse demonstrate that a hidden antiferromagnetic order
still exists, defined by a nonlocal string order
parameter(SOP)\cite{mk89}. This nonvanishing SOP was then shown be
related to the breakdown of hidden $Z_2 \times Z_2$ topological
symmetry in AKLT model\cite{kt92}. Furthermore Venuti verifies the
connection between the localized entanglement and SOP for spin-1 VBS
state\cite{venuti05}. For higher integer-spin VBS state, several
generalizations of SOP have been proposed to characterize the hidden
topological symmetry\cite{oshikawa, ts95, tu08}.

Since it is well accepted that the local operation and classical
communication(LOCC) cannot increase entanglement\cite{vidal}, the
appearance of maximal localized entanglement in AKLT model thus
means that quantum entanglement is predominant in VBS state. With
respect of this point, the bipartite entanglement for VBS state has
been first discussed\cite{fan04, katsura}, where the block
entanglement entropy is found to be saturated as $2\ln(s+1)$ for
spin-$s$ VBS state. Unfortunately the behavior of entanglement
entropy cannot present a clear explanation for the appearance of
maximal localized entanglement in VBS state. Moreover the block
entanglement entropy is insensible to the degeneracy induced by the
topological symmetry. Furthermore the dependency on the parity of
spin number $s$ is absent for the entanglement entropy of VBS state,
which has been manifested clearly by SOP\cite{ts95}. In our point
this defect could attribute to the fact that the overall
information, embedded in many-body systems, is inevitably lost due
to the trace-out of the superfluous degrees of freedom for obtaining
the reduced density matrix. And hence bipartite entanglement seems
have limited ability of the complete characterization of VBS state.

With these points the multipartite entanglement of VBS state is
discussed in this paper in order to obtain the overall information
embedded in VBS state. As shown in this paper, some distinguished
features for VBS state can be obtained from the evaluation of
multipartite entanglement, and moreover the multipartite
entanglement of VBS state demonstrates different asymptotic
behaviors from entanglement entropy. The paper is organized as the
following. In Sec.2, geometric entanglement(GE) is introduced for
the measure of multipartite entanglement. And Valence-Bond-Solid
state is defined with different boundary conditions in Sec.3. In
Sec.4, the main part of this paper, we present a detailed evaluation
of GE. Finally the conclusions and further discussion are presented
in Sec.5.

\section{Geometric Entanglement}
Geometric entanglement(GE) is used to measure the multipartite
entanglement in this paper, which is first introduced by Shimony for
bipartite pure state\cite{shimony95} and generalized to the
multipartite case by Carteret {\it et.al.}\cite{chs00}, Barnum and
Linden\cite{bl01}, Wei and Goldbart\cite{wg03}, to the mixed state
by Cao and Wang\cite{cw07}. Geometric entanglement is a genuine
multipartite entanglement measurement. The main idea of GE is to
minimize the distance $D$ between the entangled state $\ket{\Psi}$
and the fully separable state $\ket{\Phi}$ in the Hilbert space,
\begin{equation}\label{D}
D=\min_{\{\ket{\Phi}\}}\{\|\ket{\Psi}-\ket{\Phi}\|^2\}.
\end{equation}
Given the normalized $\ket{\Psi}$ and $\ket{\Phi}$, the evaluation
of $D$ is reduced to find the maximal overlap\cite{wg03}
\begin{equation}
\Lambda(\Psi)=\max_{\{\ket{\Phi}\}}|\inp{\Phi}{\Psi}|.
\end{equation}
Geometrically $\Lambda(\ket{\Psi})$ depicts the overlap angle
between the vectors $\ket{\Psi}$ and $\ket{\Phi}$ in Hilbert space.
Then the larger $\Lambda(\ket{\Psi})$ is, the shorter is the
distance and the less entangled is $\ket{\Psi}$. In order to measure
the entanglement in many-body systems, in this paper we adopt the
definition in Ref.\cite{wei05}
\begin{equation}\label{ge}
\varepsilon=\lim_{L\rightarrow\infty}-\frac{\log_2\Lambda^2(\Psi)}{L},
\end{equation}
where $L$ denotes the chain length. Eq.\eqref{ge} actually defines
the average entanglement per particle, and is also entanglement
monotone.

With this definition, the calculation for $\Lambda(\ket{\Psi})$ is
generally difficult since the optimization. Recently two works prove
independently that for the permutationally invariant entangled pure
state the maximization can be attained \emph{necessarily} by
choosing a permutationally invariant pure fully separable state at
the same time, whose amplitudes are all non-negative in a
computational basis\cite{wei09}. Moreover this conclusion has been
generalized to any symmetric pure multipartite entangled state as
shown in Ref.\cite{hkwg09}. This point could be understood properly
by noting the fact that $|\inp{\Phi}{\Psi}|=1$ means that
$\ket{\Phi}$ and $\ket{\Psi}$ depict the same state, and both of
them have the same physical features. While, $|\inp{\Phi}{\Psi}|=0$
means that $\ket{\Phi}$ and $\ket{\Psi}$ have distinct physical
features, and one can easily differentiate one from another by
physical measurements. Thus for the purpose of maximizing
$|\inp{\Phi}{\Psi}|$, it is necessary to find the fully separated
$\ket{\Phi}$ which has the same overall features to that of
$\ket{\Psi}$.

Here we present a formal proof for this intuitive speculation. The
start point for this proof is Cachy-Schwartz(CS) inequality
\begin{equation}
|\inp{v}{w}|^2\leq\inp{v}{v}\inp{w}{w},
\end{equation}
where the equality occurs if and only if the two vectors $\ket{v}$
and $\ket{w}$ in Hilbert state are linearly related, i.e.
$\ket{v}=c\ket{w}$ for some scalar $c$. It should emphasize that $c$
is \emph{unnecessary} a constant, and generally would contain other
variables. The trivial case for the equality occurring is that
$\ket{v}$ and $\ket{w}$ are the same vector aside from the
normalization constants. But for more general case, it is a
nontrivial task for given $\ket{v}$ to find a physically different
state $\ket{w}$ in order to maximize the overlap. However the task
would become apparent if one rewrites the condition for the equality
as
\begin{equation}
\ket{v}\bra{v} \ \ket{w}=\inp{v}{v}\ket{w}.
\end{equation}
This transformation implies strongly that this task is reduced to
find the eigenvector for operator $\ket{v}\bra{v}$. This task seems
trivial in the Hilbert space specified by vector $\ket{v}$. However
it becomes nontrivial for enlarged Hilbert space. Fortunately recall
that if two operators are commutative, they then share the same set
of the eigenvectors, and then $\ket{w}$ has same global symmetry to
$\ket{v}$. Consequently this task can be reduced to find the all
symmetries $T$ for operator $\ket{v}\bra{v}$. With this tricky, the
optimal process in the definition of GE can be simplified greatly.

\section{Valence-Bond-Solid state}
Dependent on the boundary conditions, integer spin-$s$ VBS state can
be defined respectively as \cite{ts95}
\begin{eqnarray}\label{pvbs}
\ket{\text{VBS}}_{\text{PBC}}=\prod_{k=1}^{L}(a_k^{\dagger}b_{k+1}^{\dagger}-b_k^{\dagger}a_{k+1}^{\dagger})^s\ket{0}
\end{eqnarray}
with periodic boundary condition(PBC)
$a^{(\dagger)}_{L+1}=a^{(\dagger)}_1,
b^{(\dagger)}_{L+1}=b^{(\dagger)}_1$, in which  $a_k^{(\dagger)},
b_k^{(\dagger)}$ are the Schwinger boson operators, $L$ is the chain
length and $\ket{0}$ denotes the vacuum state, or
\begin{eqnarray}\label{ovbs}
\ket{\text{VBS}; p, q}&=&Q_{\text{left}}(a_1^{\dagger},
b_1^{\dagger};p)
\prod_{k=1}^{L-1}(a_k^{\dagger}b_{k+1}^{\dagger}-b_k^{\dagger}a_{k+1}^{\dagger})^s
\nonumber \\&&Q_{\text{right}}(a_L^{\dagger},
b_L^{\dagger};q)\ket{0},
\end{eqnarray}
with open boundary condition(OBC)
\begin{eqnarray}\label{bc}
Q_{\text{left}}(a_1^{\dagger},
b_1^{\dagger};p)=\sqrt{_sC_{p-1}}(a_1^{\dagger})^{s-p+1}(b_1^{\dagger})^{p-1}
\nonumber\\
Q_{\text{right}}(a_L^{\dagger},
b_L^{\dagger};q)=\sqrt{_sC_{q-1}}(a_L^{\dagger})^{q-1}(b_L^{\dagger})^{s-q+1}.
\end{eqnarray}
in which $p,q=1, \cdots, s+1$ and ${_{n}C_{m}}$ denotes the binomial
function. For OBC the VBS state is $(s+1)^2$-fold degenerate.

VBS state can be rewritten as the matrix-product form\cite{ts95}
\begin{eqnarray}\label{mat}
\ket{\text{VBS}}_{\text{PBC}}&=&\text{Tr}[g_1\cdot g_2\cdots g_L]\nonumber\\
\ket{\text{VBS};p,q}&=&[g_{\text{start}}\cdot g_2\cdots g_L]_{(p,q)}
\end{eqnarray}
where $(p, q)$ denote the coordinates of matrix elements, and the
matrix $g_i$ is $(s+1)\times (s+1)$ dimension, of which the element
reads
\begin{eqnarray}\label{g}
g_i(p,q)&=&(-1)^{s+p-1}\sqrt{{_{s}C_{p-1}} {_{s}C_{q-1}}}\sqrt{(s-p+q)!(s+p-q)!}\ket{s;q-p}_i\nonumber\\
g_{\text{start}}(p,q)&=&\sqrt{{_sC_{p-1}}
{_sC_{q-1}}}\sqrt{(s-p+q)!(s+p-q)!}\ket{s;q-p}
\end{eqnarray}
Obviously $g_i$ is related only to the $i$-th spin. Manifestly VBS
state with PBC is permutationally invariant, while it is not so for
OBC because of the open ends.

\textit{-Symmetry-} VBS state is the ground state of Hamiltonian
\begin{equation}\label{ham}
H_{\text{VBS}}=\sum_{i=1}^{L}P_{s+1}^{2s}(\mathbf{S}_i\cdot\mathbf{S}_{i+1}),
\end{equation}
where the operator $P_{s+1}^{2s}$ denotes to project the spins at
sites $i$ and $i+1$ into the subspace with the total spin $J=s+1,
\cdots, 2s$ \cite{aklt, arovas88}. Since Hamiltonian Eq.\eqref{ham}
is SU(2) invariant, then $[H, \prod_{i=1}^{L}\exp(i\pi S_i^z)]=0$.
This symmetry is also required for VBS state
\begin{equation}\label{symmetry}
[\ket{\text{VBS}}\bra{\text{VBS}}, \prod_{i=1}^{L}\exp(i\pi
S_i^z)]=0,
\end{equation}
where $\ket{\text{VBS}}$ denotes the VBS state without specifying
the boundary conditions.

\textit{-Magnetization-} For PBC,  each $S_i^z$ eigenvalue occurs
with equal probability\cite{ts95}, and then the reduced density
matrix of single spin is the unit matrix. This implies that each
eigenvalue of the total $S_z=\sum_{i=1}^L S_i^z$ also occurs with
equal probability, and the total magnetization $\langle S_z
\rangle_{VBS}$ in this case is zero. Furthermore $\langle S_y
\rangle_{VBS}=\langle S_x \rangle_{VBS}=0$. Thus VBS state for PBC
is rotationally invariant.

The situation becomes different for OBC. VBS is not necessarily
rotationally invariant because of the open ends. Moreover the open
ends may induce a local perturbation to the spin chain, and the
hidden topological property for VBS state can be obtained from the
system's responds to this local perturbation.

\section{GE in VBS state}
Given these properties, one can now determine the fully separable
state $\ket{\Phi}=\otimes_{i=1}^{L}\ket{\phi _i}$, in which
$\ket{\phi _i}=\sum_{m=-s}^sc_m^{(i)}\ket{m}_i$ on the basis of
$\{S^{(i)}_z;\ket{m}_i\}$ and the coefficient $c_m^{(i)}$ is
generally complex and dependent on the position $i$. With the
requirement of Eq.\eqref{symmetry}, one has $[\ket{\Phi}\bra{\Phi},
\prod_{i=1}^{L}\exp(i\pi S^{(i)}_z)]=0$. Then two different
situations can be identified as
\begin{eqnarray}
\exp(i\pi S^{(i)}_z)\ket{\phi _i}^p&=&\ket{\phi _i}^p\nonumber\\
\exp(i\pi S^{(i)}_z)\ket{\phi _i}^n&=&-\ket{\phi _i}^n
\end{eqnarray}
where
\begin{eqnarray}
\ket{\phi _i}^p&=&\sum_{k=-[s/2]}^{[s/2]}c^{(i)}_{k}\ket{2k}_i\nonumber\\
\ket{\phi
_i}^n&=&\sum_{k=-[(s+1)/2]}^{[(s-1)/2]}c^{(i)}_{k}\ket{2k+1}_i
\end{eqnarray}
and $[n/2]$ denotes the maximal integer number not bigger than
$n/2$. Then one has to find the maximal overlap among
$\{|\bra{\text{VBS}}\otimes_{i=1}^{L}\ket{\phi _i}^p|,
|\bra{\text{VBS}}\otimes_{i=1}^{L}\ket{\phi _i}^n|\}$. Consequently
GE is determined by the minimization
\begin{equation}
\varepsilon=\min\{\varepsilon_p, \varepsilon_n\}
\end{equation}
where $\varepsilon_{p(n)}$ denotes GE evaluated with the separable
state $\otimes_{i=1}^{L}\ket{\phi}^{p(n)}_i$ respectively. Dependent
on the boundary conditions, the following discussion has to be
divided into two subsections.

\subsection{Periodic Boundary Condition}
\begin{figure}[t]
\center
\includegraphics[bbllx=16, bblly=16, bburx=271, bbury=185,width=5.5cm]{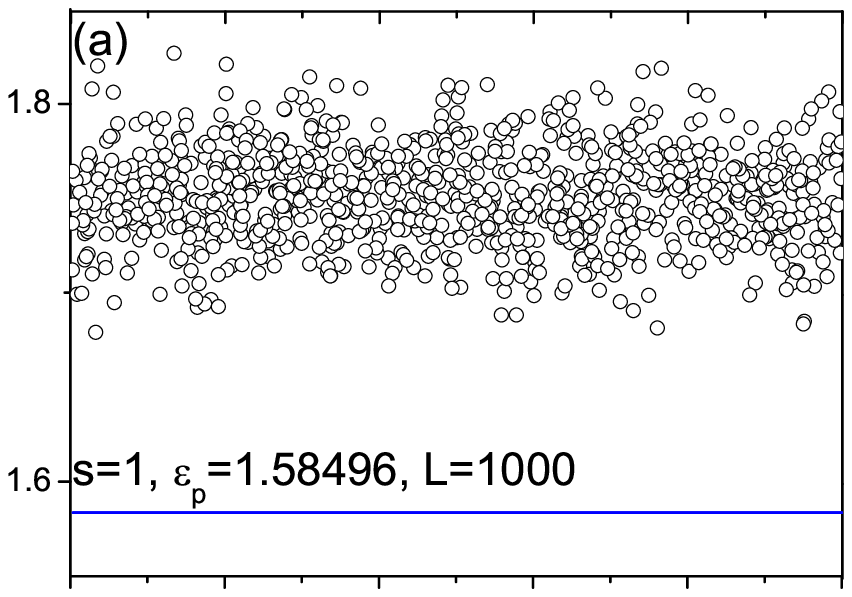}
\includegraphics[bbllx=16, bblly=16, bburx=271, bbury=185,width=5.5cm]{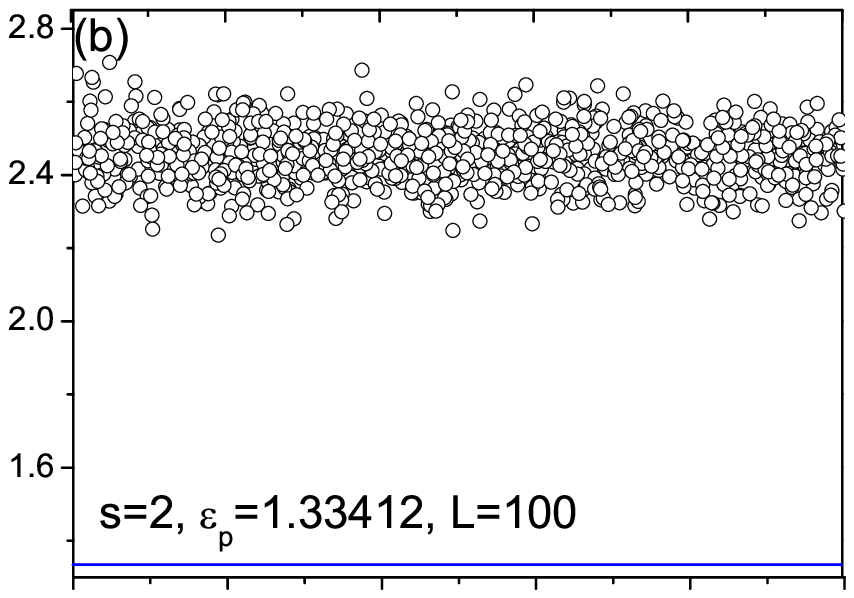}
\includegraphics[bbllx=16, bblly=16, bburx=271, bbury=185,width=5.5cm]{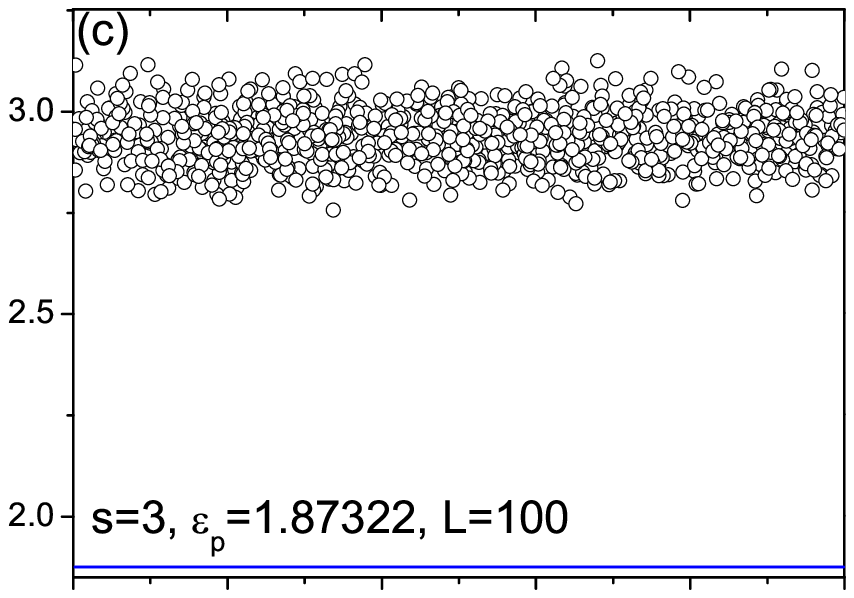}
\includegraphics[bbllx=16, bblly=16, bburx=264, bbury=185,width=5.5cm]{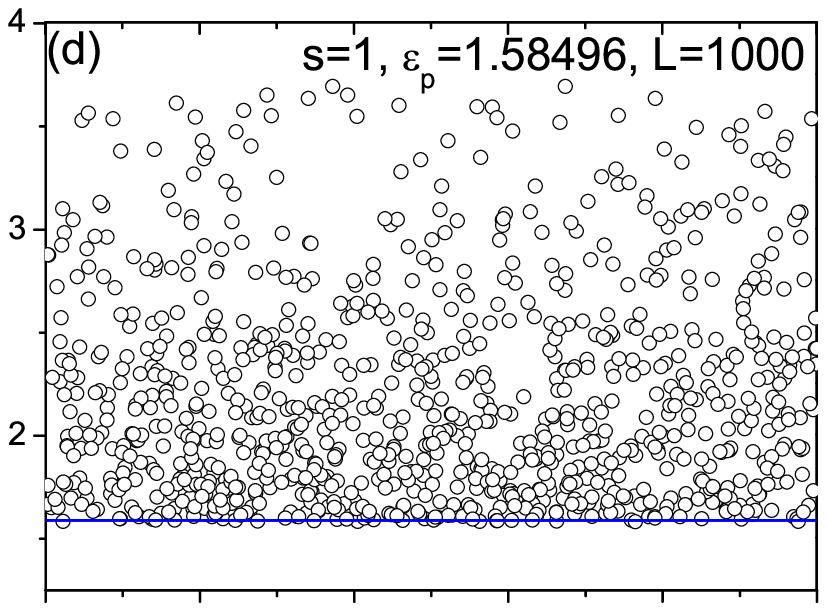}
\includegraphics[bbllx=16, bblly=16, bburx=266, bbury=185,width=5.5cm]{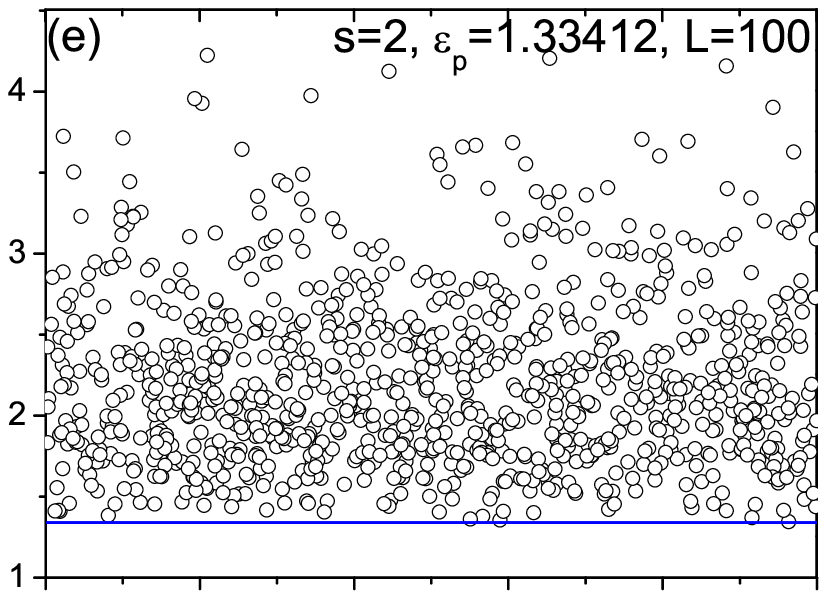}
\includegraphics[bbllx=16, bblly=16, bburx=266, bbury=185,width=5.5cm]{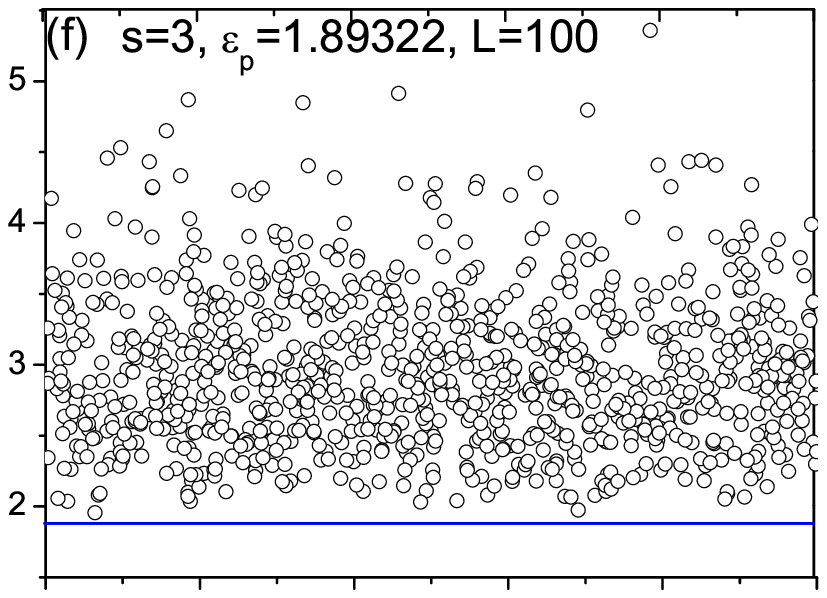}
\caption{\label{npbc}(Color Online)The randomly numerical simulation
of $-\tfrac{\log_2 |\inp{\Phi}{\text{VBS}}|^2}{L}$ for VBS state
with PBC and spin number $s=1, 2, 3$ respectively. There is 1000
sample data for every simulation. The plots of (a), (b) and (c)
correspond to the simulations without any other restriction on the
coefficients $c_k^{(i)}$ of the fully separable state $\ket{\Phi}$
except of the requirement of normalization. While for plots (d), (e)
and (f) the requirement of permutational invariance for $\ket{\Phi}$
has been implemented, but $c_k$ for different $k$ is still chosen
randomly. The (blue) solid line corresponds the value of GE after
applying the all approximations stated in the paper. }
\end{figure}

As displayed in Sec.3, VBS state is permutationally invariant in
this case. It has been proved in Refs. \cite{wei09,hkwg09} that the
fully separable state is necessarily also permutationally invariant
in order to maximize the overlap. Then the fully separable state
$\ket{\Phi}$ becomes in this case
\begin{eqnarray}
\ket{\Phi}_p&=&(\ket{\phi}^p)^{\otimes L}\nonumber\\
\ket{\Phi}_n&=&(\ket{\phi}^n)^{\otimes L}.
\end{eqnarray}

Another important feature is that VBS state is rotationally
invariant for PBC, and hence has zero total spin\cite{arovas88}. It
is natural to endow the fully separable state $\ket{\Phi}_{p(n)}$
with the same character. This point is evident to note that it
belongs to completely different space for the states which has
vanishing total spin or not, and then the overlap between states in
the two distinct spaces must be zero.

With this requirement, the coefficient $c^{(i)}_k$ preferably has
the same amplitude, independent both of the position $i$ and quantum
number $k$. Moreover $c^{(i)}_k$ should preferably be in phase in
order to maximize the overlap with respect of the special form of
VBS state shown by Eq.\eqref{pvbs}. This point becomes clear by
noting first that ${^p\bra{\phi}}\otimes g_i$ has a block-diagonal
form dependent on the parity of $p,q$ for any spin $s$ (see Appendix
A for more detail). Set $c_k=c_{p-q}=|c_{p-q}|e^{i\theta_{p-q}}$, of
which $\theta_{q-p}$ and $|c_{p-q}|$ depend only on the difference
$|q-p|$. And then
\begin{eqnarray}
&&{^p\bra{\phi}}\otimes g_i= \nonumber\\ &&\left(
\begin{array}{cc} \dashbox(130,20){$|c_{p-q}|e^{i\theta_{q-p}}g_{p, q}$(even $p$ and $q$)} & 0 \\ 0 &
\dashbox(125,20){$|c_{p-q}|e^{i\theta_{q-p}}g_{p, q}$(odd $p$ and
$q$)}\end{array}\right)
\end{eqnarray}
where the dashed boxes denote the sub-matrixes satisfying the
conditions stated in the brackets, and the element of the sub-matrix
is read
\begin{eqnarray*}
g_{p,q}=(-1)^{s+p-1}\sqrt{{_{s}C_{p-1}}{_{s}C_{q-1}}}\sqrt{(s-p+q)!(s+p-q)!}.
\end{eqnarray*}
Then one has
\begin{eqnarray}
|{_p\inp{\Phi}{VBS}_{\text{PBC}}}|&=&\left|\text{Tr}[{_p\bra{\Phi}}g_1\cdot
g_2\cdots g_L]\right|=\left|\sum^{s+1}_{p_1,p_2, \cdots p_L; q_1,
q_2, \cdots,
q_L=1 }\prod_{i=1}^{L}|c_{p_i-q_i}|\prod_{i=1}^{L}e^{i\theta_{q_i-p_i}}g_{p_i, q_i}\right|\nonumber\\
&\leq&|c_p|^L\left|\sum^{s+1}_{p_1,p_2, \cdots p_L; q_1, q_2,
\cdots, q_L=1 }\prod_{i=1}^{L}e^{i\theta_{q_i-p_i}}g_{p_i,
q_i}\right|\nonumber\\
&\leq&|c_p|^L\sum^{s+1}_{p_1,p_2, \cdots p_L; q_1, q_2, \cdots,
q_L=1 }\prod_{i=1}^{L}|e^{i\theta_{q_i-p_i}}g_{p_i, q_i}|\nonumber\\
&=&|c_p|^L\sum^{s+1}_{p_1,p_2, \cdots p_L; q_1, q_2, \cdots, q_L=1
}\prod_{i=1}^{L}|g_{p_i, q_i}|,
\end{eqnarray}
where the following relation is used
\begin{eqnarray}
\prod_{i=1}^{L}|c_{p_i-q_i}|\leq\frac{\sum_{i=1}^L
|c_{p_i-q_i}|^L}{L},
\end{eqnarray}
in which the equality occurs if and only if
$|c_{p_i-q_i}|=|c_{p_j-q_j}|=|c_p|$ for arbitrary $i\neq j$.

With even $L$ and noting the common factor $(-1)^{s+p-1}$ in $g_{p,
q}$, then
\begin{equation}
|{_p\inp{\Phi}{VBS}_{\text{PBC}}}|\leq |c_p|^L\sum^{s+1}_{p_1,p_2,
\cdots p_L; q_1, q_2, \cdots, q_L=1 }\prod_{i=1}^{L}g_{p_i, q_i}.
\end{equation}
which means that one need only choose $c_p$ being non-negative real
for the maximization of the overlap. As for odd $L$, one should note
that ${^p\bra{\phi}}\otimes g_i$ is traceless and the overlap always
vanishes.

For $\ket{\phi}^n$, ${^n\bra{\phi}}\otimes g_i$ has a
block-anti-diagonal form with opposite sign for the two blocks, as
shown in Appendix A, and thus the overlap always vanishes for odd
$L$. For even $L$ one only notes that the product of two
block-anti-diagonal matrixes is block-diagonal, and the two diagonal
block has the same sign after this product. Then the analysis is
applicable in this case.

Then the fully separable state can be determined exactly
\begin{eqnarray}\label{phi}
\ket{\Phi}_p&=&\bigotimes_{i=1}^{L}c_p\sum_{k=-[s/2]}^{[s/2]}\ket{2k}_i\nonumber\\
\ket{\Phi}_n&=&\bigotimes_{i=1}^{L}c_n\sum_{k=-[(s+1)/2]}^{[(s-1)/2]}\ket{2k+1}_i
\end{eqnarray}
in which $c_p=1/\sqrt{1+2[s/2]}, c_n=1/\sqrt{2[(s+1)/2]}$. Then with
respect of the matrix-product form of VBS state, the overlap can be
expressed readily as
\begin{eqnarray}
\Lambda_p&=&\frac{|{_p\inp{\Phi}{VBS}_{\text{PBC}}}|}{\sqrt{_{\text{PBC}}\inp{VBS}{VBS}_{\text{PBC}}}}
=\frac{\text{Tr}[{_p\bra{\Phi}}g_1\cdot g_2\cdots
g_L]}{\sqrt{\text{Tr}[g^{*}_1\cdot g^{*}_2\cdots
g^{*}_L\otimes g_1\cdot g_2\cdots g_L]}}\nonumber\\
\Lambda_n&=&\frac{|{_n\inp{\Phi}{VBS}_{\text{PBC}}}|}{\sqrt{_{\text{PBC}}\inp{VBS}{VBS}_{\text{PBC}}}}
=\frac{\text{Tr}[{_n\bra{\Phi}}g_1\cdot g_2\cdots
g_L]}{\sqrt{\text{Tr}[g^{*}_1\cdot g^{*}_2\cdots g^{*}_L\otimes
g_1\cdot g_2\cdots g_L]}},
\end{eqnarray}
where $g^{*}$ denotes to take Hermitian conjugate for the matrix
elements, but not transpose the matrix.

In order to demonstrate more explicitly the validity of the
statements above, the randomly numerical simulation has been
implemented for several simple cases, as shown in Fig.\ref{npbc}.
For Fig.\ref{npbc}(a), (b) and (c), there is no any other
restriction on the choice of $c_k^{(i)}$ except of the requirement
of normalization. It is obvious that the sample data are always
bigger than the value of GE evaluated after applying the
approximations. Furthermore the approximation that $c^{(i)}_k$ is a
positive constant independent both of the coordinate $i$ and quantum
number $k$, has also been verified numerically as shown in
Fig.\ref{npbc}(d), (e) and (f). The results show that this
approximation is really proper and correct. With the assistance of
the numerical simulation, GE can be evaluated readily.

\begin{figure}[t]
\center
\includegraphics[bbllx=48, bblly=44, bburx=800, bbury=251,width=17cm]{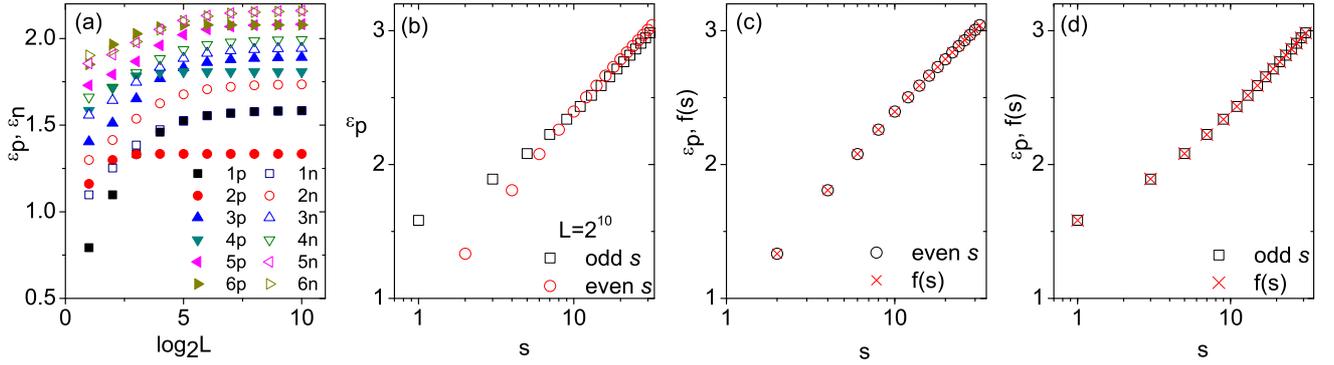}
\caption{\label{pbc}(Color Online)(a) $\varepsilon_p, \varepsilon_n$
versus the particle number $L$ with PBC. In this figure the Arab
numbers denote the spin and $p,n$ denote the plotting for
$\varepsilon_p, \varepsilon_n$ respectively; (b) the scaling of
$\varepsilon_p$ with respect to spin $s$ and the  fitting function
$f(s)$; Use $f(s)$ to fit GE for even(c)and odd(d) spin-s case.}
\end{figure}

For arbitrary spin-$s$ the analytical expressions for the overlap
are difficult to obtain. Thus $\varepsilon_p, \varepsilon_n$ are
plotted numerically in Fig. \ref{pbc}(a). It is obvious that
$\varepsilon_p, \varepsilon_n$ tend to be saturated with the
increment of particle number $L$. Moreover one can note
$\varepsilon_p<\varepsilon_n$ under large $L$. As our study is
implemented mainly under the limit of large $L$, the discussions
below would focus on $\varepsilon_p$.

It should emphasize that since $\varepsilon_{p(n)}$ denotes the
average GE per particle, the saturation for $\varepsilon_{p(n)}$
shown in Fig. \ref{pbc}(a) means that the globe GE $E_G$ for VBS
state, defined first in Ref.\cite{odv08}, should be proportional to
the particle number $L$, and is divergent under
$L\rightarrow\infty$, i.e.
\begin{equation}
E_G=\varepsilon_p L.
\end{equation}
This phenomena is distinct from the entanglement entropy of VBS
state, which is shown to be saturated as $2\ln(s+1)$ because of the
appearance of energy gap above the ground state\cite{fan04,
katsura}. However the divergency of $E_G$ strongly implies that
there is hidden degeneracy not captured by entanglement entropy.
With the findings of the breakdown of hidden $Z_{s+1}\times Z_{s+1}$
topological symmetry\cite{kt92,tu08}, it hints that GE would be also
sensible to the existence of topological phases in many-body
systems. A further discussion would be presented in the final
section.

Furthermore, as shown in Fig.\ref{pbc}(b) where the scaling behavior
of $\varepsilon_p$ with spin-$s$ is plotted, $\varepsilon_p$ shows
two different scaling behavior, dependent on even or odd spin $s$.
This feature is consistent with the observation that one have to
define different SOPs for odd or even $s$ VBS state. A fitting
function $f(s)$ can be constructed for identifying the scaling
behavior of GE with spin $s$,
\begin{equation}\label{f1}
f(s)=\alpha\log (s+\frac{\beta}{s}+\gamma)+\delta
\end{equation}
in which $\alpha, \beta, \gamma$ and $\sigma$ are tunable
parameters. The chosen values of these parameters for the fittings
of $\varepsilon_p$ are listed in Table.\ref{f2}. It has been shown
that the scaling behavior with spin $s$ for entanglement entropy of
VBS state\cite{fan04, katsura} is attributed to the tendency to be
unit form for the reduced block density matrix under limit
$L\rightarrow \infty$\cite{katsura}. Comparably it is unclear until
now for the distinct feature of GE disclosed by Eq.\eqref{f1}.

The differences between entanglement entropy and GE in VBS state can
intuitively attribute to the trace-out of the superfluous degrees of
freedom for obtaining the reduced density matrix in order to
calculate the block entanglement entropy. Thus the overall
information about VBS state is lost because of this trace-out. With
respect of this point, it seems not enough by only sampling a
portion of the system to obtain the overall information of many-body
systems.

\begin{table}\label{f2}
\center \caption{The chosen parameters for the fittings of GE in
Figs.\ref{pbc}(c) and (d).}
\begin{tabular}{|c|c|c|c|c|}
\hline
  & $\alpha$ & $\beta$ & $\gamma$ & $\delta$\\
\hline
Fig. \ref{pbc}(c) & 1.41 & -0.39 & 2.82 & 0.83 \\
Fig. \ref{pbc}(d) & 1.15 & 1.67 & -1.83 & 1.33\\
 \hline
\end{tabular}
\end{table}

Another interesting character for this numerical evaluation is that
$\varepsilon_p, \varepsilon_n$ are sensitive to the parity of
particle number $L$. As for odd spin $s$ case, $\Lambda_p$ is always
vanishing for odd $L$ and then $\varepsilon_p$ is infinity. This
point can be easily verified (see Appendix A). Whereas for even
spin-$s$ our calculation shows that  $\Lambda_p$ for odd or even $L$
tends to be consistent numerically with the increment of particle
number. The similar behavior also happens for the evaluation of
$\varepsilon_n$.

\subsection{Open Boundary Condition}

\begin{figure}[t]
\center
\includegraphics[bbllx=16, bblly=16, bburx=271, bbury=185,width=5.5cm]{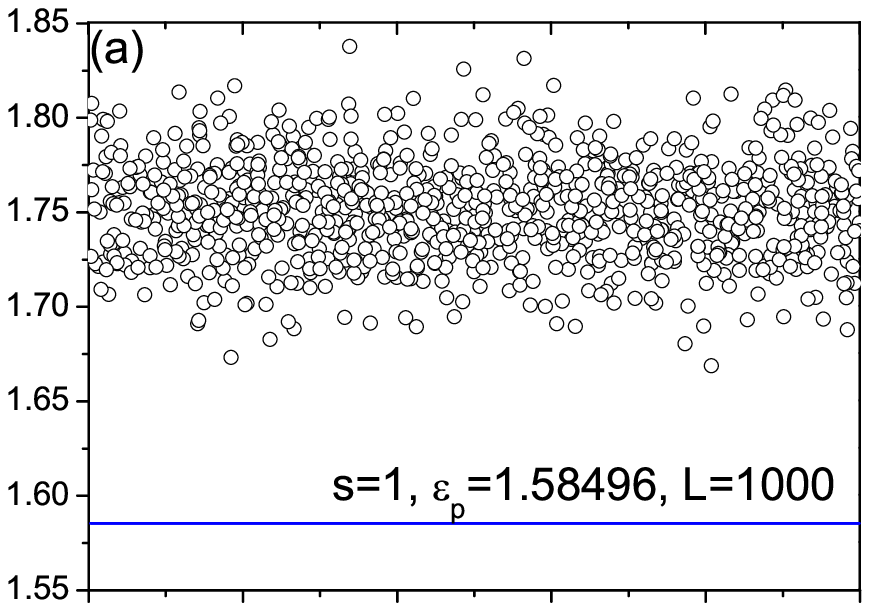}
\includegraphics[bbllx=16, bblly=16, bburx=271, bbury=185,width=5.5cm]{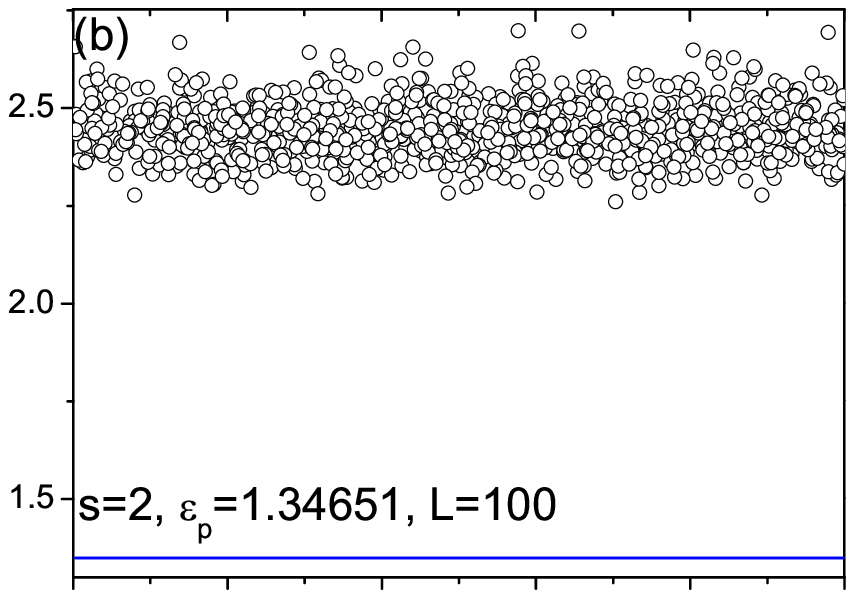}
\includegraphics[bbllx=16, bblly=16, bburx=271, bbury=185,width=5.5cm]{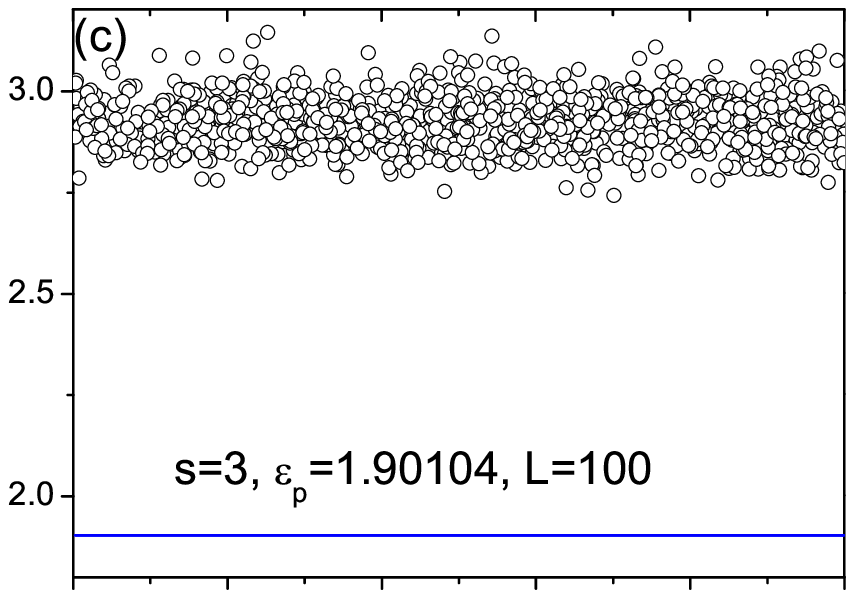}
\includegraphics[bbllx=16, bblly=16, bburx=264, bbury=185,width=5.5cm]{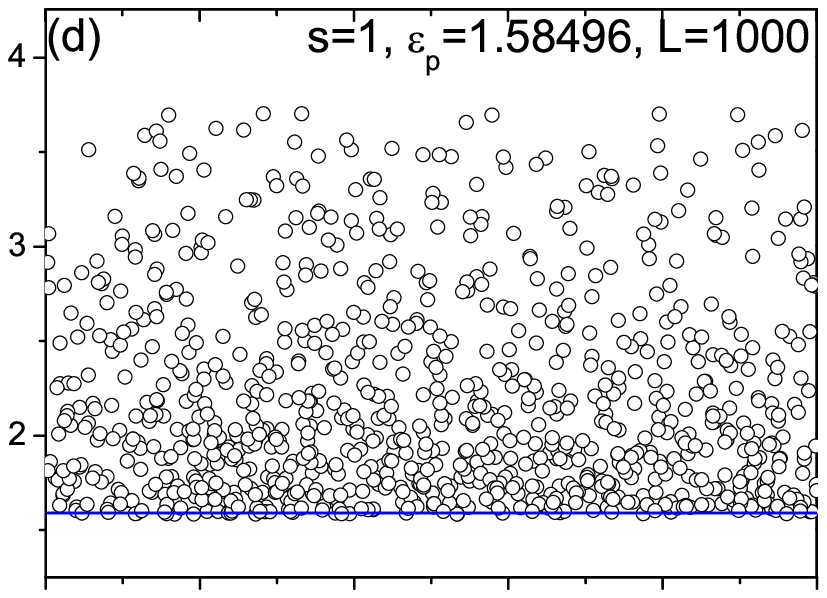}
\includegraphics[bbllx=16, bblly=16, bburx=266, bbury=185,width=5.5cm]{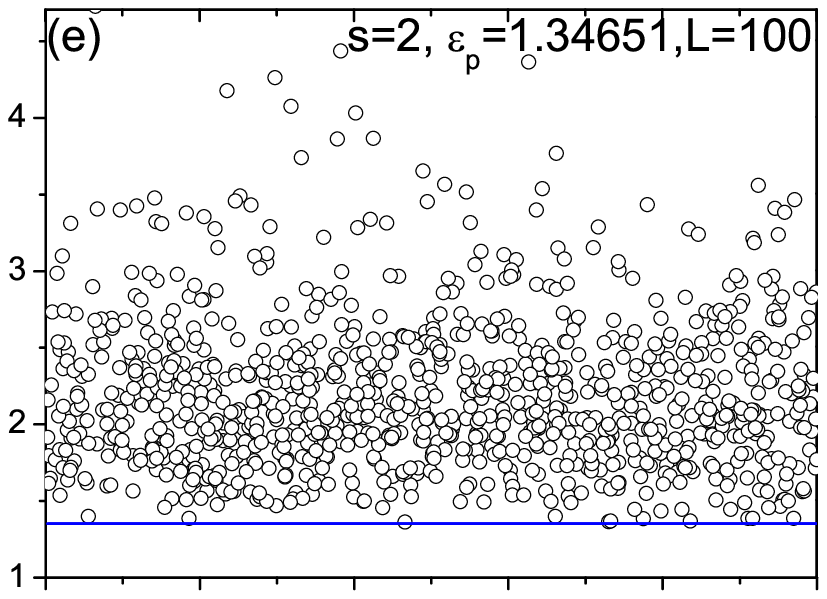}
\includegraphics[bbllx=16, bblly=16, bburx=266, bbury=185,width=5.5cm]{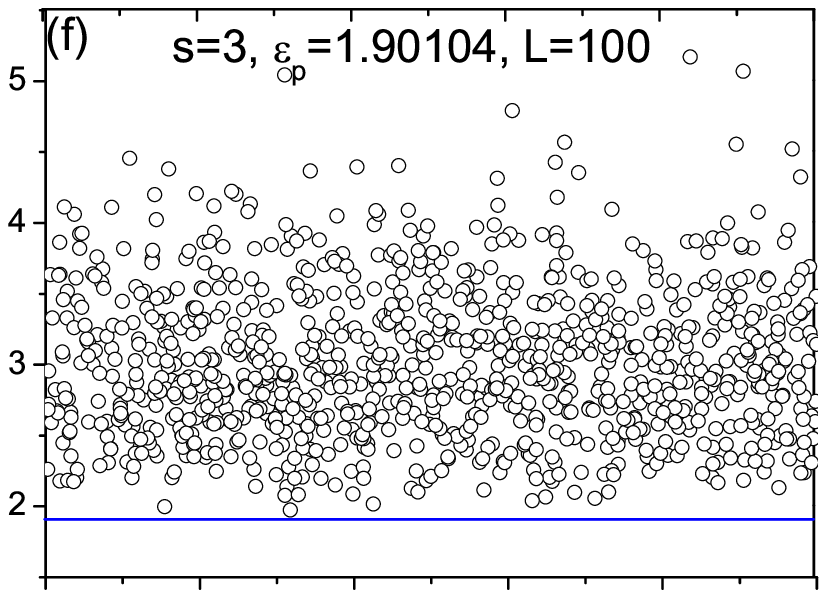}
\caption{\label{nobc}(Color Online)Color Online)The randomly
numerical simulation of $-\tfrac{\log_2
|\inp{\Phi}{\text{VBS}}|^2}{L}$ for VBS state with OBC and spin
number $s=1, 2, 3$ respectively. There is 1000 sample data for every
simulation. The plots of (a), (b) and (c) correspond to the
simulations without any other restriction on the coefficients
$c_k^{(i)}$ of the fully separable state $\ket{\Phi}$ except of the
requirement of normalization. While for plots (d), (e) and (f) the
requirement of permutational invariance for $\ket{\Phi}$ has been
implemented except of the end $g_{\text{start}}$, but $c_k$ for
different $k$, including that for $g_{\text{start}}$, is still
chosen randomly. The (blue) solid line corresponds the value of GE
after applying the all approximations stated in the paper.  }
\end{figure}

The situation becomes complex for OBC. Although the requirement Eq.
\eqref{symmetry} persists, the permutational invariance for VBS is
broken in this case. Thus the assumptions made in the preceding
subsection has to be reconsidered seriously.

The crucial step is still to determine the overlap
$|\inp{\text{VBS};p,q}{\Phi}|$. It is interesting to note from the
matrix-product form of $\ket{\text{VBS};p, q}$ that except of the
end $g_{\text{start}}$ the remaining $g_i$s of this expression are
obviously permutational invariant. With respect of the findings in
Ref.\cite{wei09}, one could reasonably assume
\begin{eqnarray}
\inp{\text{VBS};p,q}{\Phi}=[{_{\text{start}}\bra{\phi'}}g_{\text{start}}{_1\bra{\phi}}g_1\cdots
{_{L-1}\bra{\phi}}g_{L-1}]_{(p,q)}
\end{eqnarray}
Now it is crucial to find the relation between $\ket{\phi'}$ and
$\ket{\phi}$.

As stated in Ref. \cite{fm91}, the boundary effect vanishes
exponentially with increasing length of the system if one focuses on
the local operator. From the definition of GE Eq.\eqref{ge}, it
actually imposes a locally independent operator
$\ket{\phi_i}\bra{\phi_i}$ on the individual spin since the choice
of $\ket{\phi_i}$ is independent with each other in the definition.
Thus it is an acceptable approximation to ignore the effect of
$g_{\text{start}}$ on determining $\ket{\phi'}$ under the limit of
large $L$, and let $\ket{\phi'}=\ket{\phi}$. Another important
character for $g_{\text{start}}\cdot g_2\cdots g_L$ is that the
eigenvectors of single spin $S_j^z$ occur with equal
probability\cite{ts95}, which also means that the different boundary
condition represented by $(p,q)$, happens with the same probability.
With these observations, it also is a reasonable assumption that the
coefficient $c_k^{(i)}$ could be same to that for PBC. Hence for OBC
the fully separable state is same as that for PBC (see Eq.
\eqref{phi}).

\begin{figure}[t]
\center
\includegraphics[bbllx=18, bblly=20, bburx=271, bbury=205,width=5.5cm]{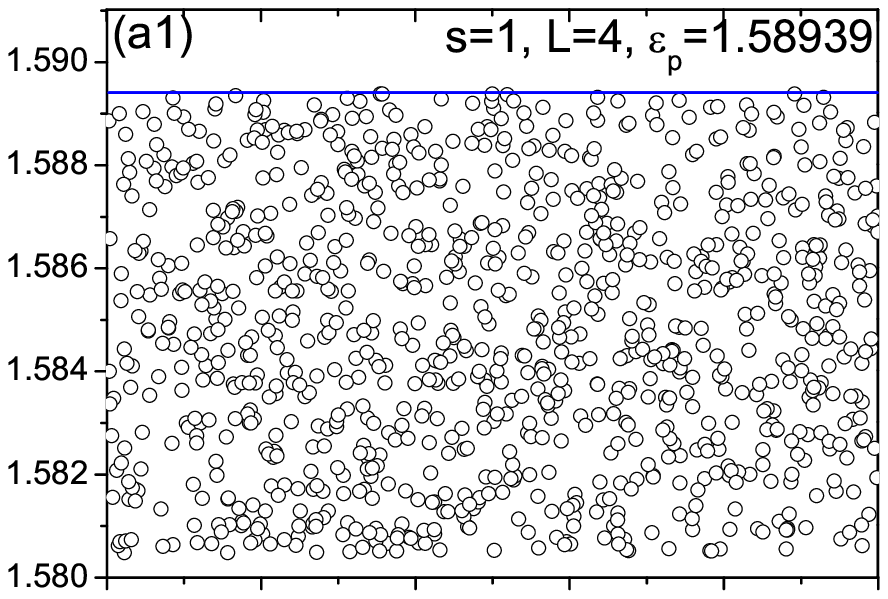}
\includegraphics[bbllx=18, bblly=20, bburx=271, bbury=205,width=5.5cm]{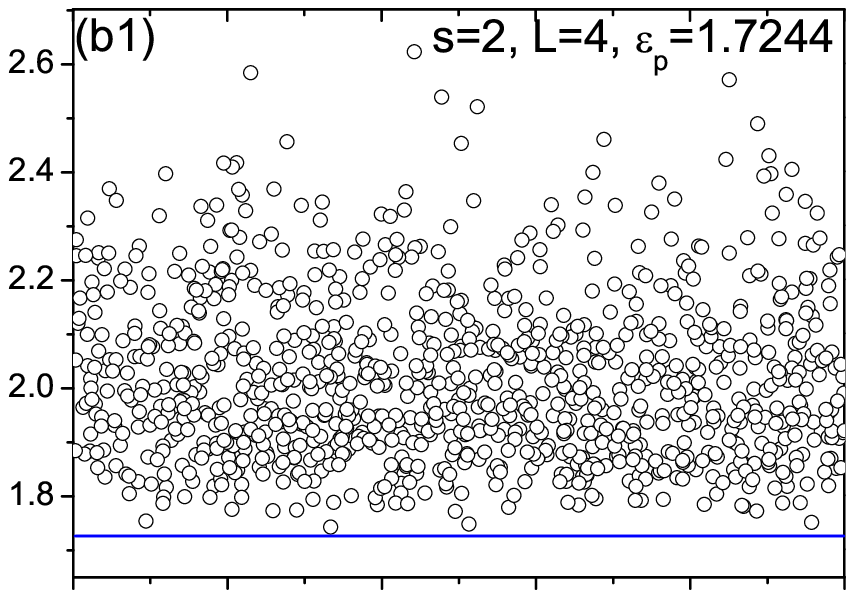}
\includegraphics[bbllx=18, bblly=20, bburx=271, bbury=205,width=5.5cm]{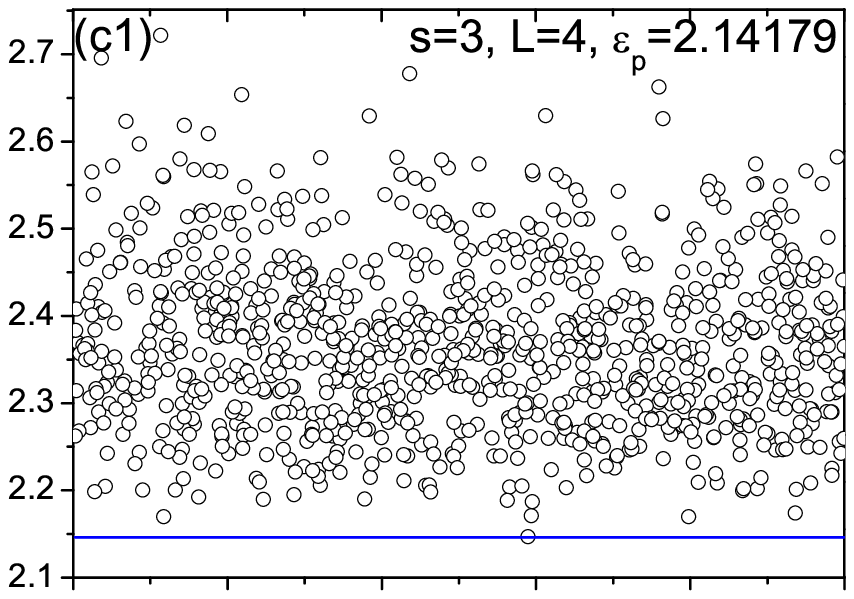}
\includegraphics[bbllx=38, bblly=17, bburx=299, bbury=200,width=5.5cm]{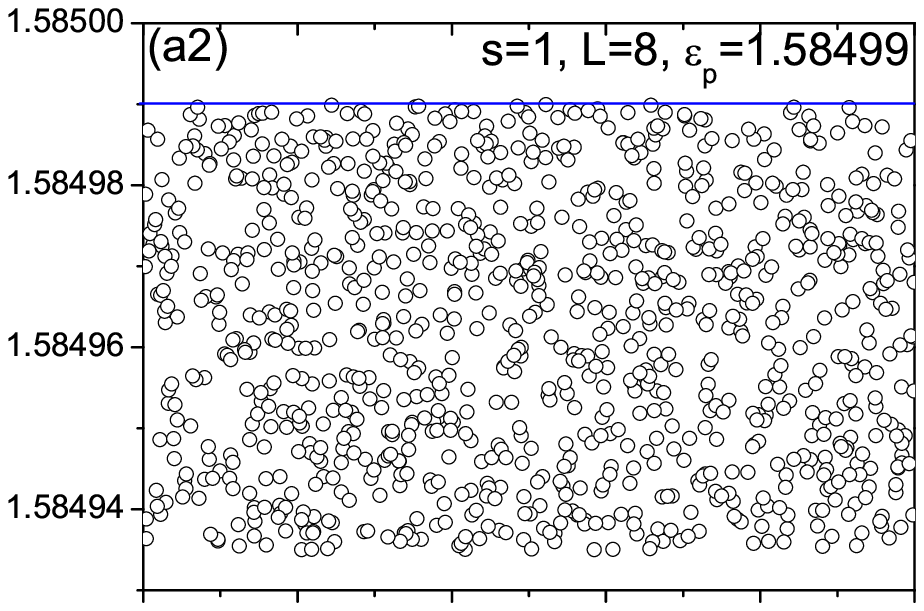}
\includegraphics[bbllx=18, bblly=20, bburx=258, bbury=205,width=5.5cm]{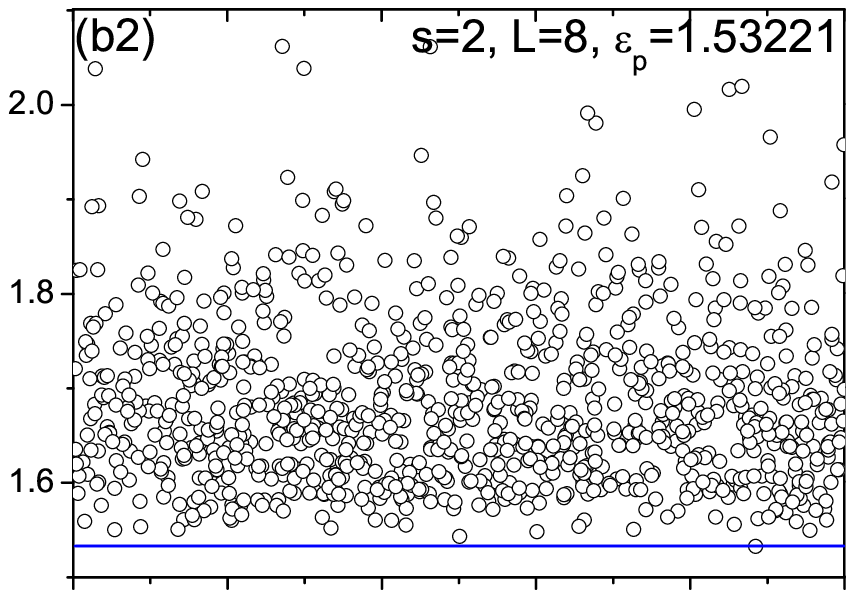}
\includegraphics[bbllx=18, bblly=35, bburx=264, bbury=219,width=5.5cm]{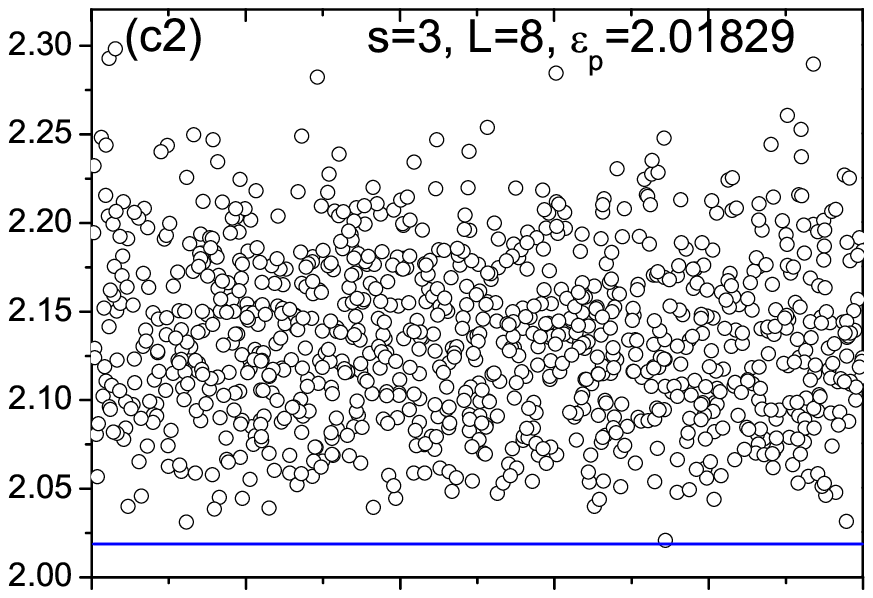}
\caption{\label{ebc}(Color Online) The numerical simulations for the
effect of OBC with different $L$. Except for the end
$g_{\text{start}}$, the state $\ket{\phi}$ is chosen based on the
approximations stated in the paper, and $\ket{\phi}_{\text{start}}$
is still randomly chosen. There is 1000 sample data for every
simulation.}
\end{figure}

\begin{figure}[t]
\center
\includegraphics[bbllx=39, bblly=35, bburx=541, bbury=241,width=11cm]{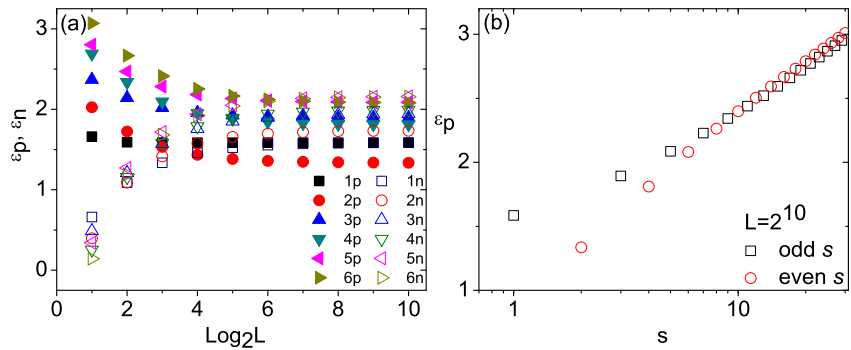}
\caption{\label{obc}(Color Online)(a) $\varepsilon_p, \varepsilon_n$
versus the particle number $L$ with OBC. In this figure the Arab
numbers denote the spin and $p,n$ denote the plotting for
$\varepsilon_p, \varepsilon_n$ respectively. (b) the scaling of
$\varepsilon_p$ with the spin $s$, which is very similar to
Fig.\ref{pbc}(b).}
\end{figure}

In order to demonstrate the validity of these approximations above,
the numerical simulations are implemented by randomly choosing
$c_k^{(i)}$. Similar to the simulation for PBC, the plots
Fig.\ref{nobc}(a), (b) and (c) display the results of simulation
without any other restriction on $c_k^{(i)}$ except of the
requirement of normalization. For plots Fig.\ref{nobc}(d), (e) and
(f), the permutational invariance is imposed on the other particles
except of the end $g_{\text{start}}$, in order to verify the
approximation that $c_k^{(i)}$ is a positive constant independent of
the coordinate $i$ and quantum number $k$. These plots manifestly
display the validity of this assumption. Furthermore the effect of
OBC has been considered independently, as shown numerically in
Fig.\ref{ebc}. For this simulation, the approximations stated in the
paragraph above are implemented for the particles except of the end,
in order to highlight the effect of boundary condition. Obviously
with the increment of particle number $L$, the effect of OBC tends
to be vanishing. And so except for $s=1$, this observation means
that these approximations are valid even for very small $L$. As for
$s=1$, our further numerical simulation shows that with the
increment of $L$, all data would become consistent with the result
obtained by the approximations above. It should point out that with
the consideration of the small $L$ $\varepsilon=\varepsilon_n$,
which is obviously below $\varepsilon_p$ as shown in Fig.\ref{obc}.
For the convenience of the comparison with previous observations,
only $\varepsilon_p$ is plotted in Fig.\ref{ebc}(a1) and (a2).

With these approximations, it is not difficult to find the overlap
$|\inp{\text{VBS};p,q}{\Phi}|$. In order to eliminate the dependence
on special boundary condition, the average for all degenerate states
is adopted
\begin{equation}
|\Lambda(\ket{\text{VBS}})|^2=\frac{1}{(s+1)^2}\sum_{(p,q)}\frac{|\inp{\text{VBS};p,q}{\Phi}|^2}{\inp{\text{VBS};p,q}{\text{VBS};p,
q}},
\end{equation}
with
\begin{equation}
\inp{\text{VBS};p,q}{\Phi}^2=[g_{\text{start}}\ket{\phi}\prod_i^{L-1}
g_i\ket{\phi}_i]_{(p,q)}^2.
\end{equation}
Since the elements of $g_i$ and $\ket{\phi}_i$ are real, one
obviously obtain under the limit of large $L$
\begin{equation}\label{ove}
|\Lambda(\ket{\text{VBS}})|^2=\frac{1}{c_L}\text{Tr}^{(2)}[g_{\text{start}}\ket{\phi}\prod_{i=1}^{L-1}
 g_i\ket{\phi}_i].
\end{equation}
where $\text{Tr}^{(2)}[M]$ denotes the sum of the square of all the
elements in matrix $M$, and the normalization constant
$c_{L}\approx(s+1)[\frac{(2s+1)!}{s+1}]^L$ for large $L$\cite{ts95}.

For $s=1$ one has $|\Lambda(\ket{\text{VBS}})|^2=3^{-L}$ for
separate states $\ket{\Phi}_p$ and $\ket{\Phi}_n$. Then we have
$\varepsilon_p=\varepsilon_n=\log_2 3$ under $L\rightarrow \infty$.
In a recent work of Or\'{u}s \cite{orus08}, the author discussed the
block GE for spin-1 VBS state. Our result is consistent with
Or\'{u}s' result when the block has only one particle.  For $s=2$,
$|\Lambda(\ket{\text{VBS}})|^2=[(4/3)^L
(1+\sqrt{6})^L+(1-\sqrt{6})^L+4^L]/c_L$ for $\ket{\Phi}_p$, and
$3^L/c_L$ for separate state $\ket{\Phi}_n$. For higher $s$, the
exact expressions for $\Lambda(\ket{\text{VBS}})$ is difficult to
find. Hence our discussion below relies on the numerical evaluation.

In Fig.\ref{obc} $\varepsilon_p, \varepsilon_n$ are plotted
respectively. As shown in Fig.\ref{obc}(a), both $\varepsilon_p,
\varepsilon_n$ tend to be saturated with the increment of $L$,
similar to the case of PBC. Moreover $\varepsilon_p$ and
$\varepsilon_n$ are also generally unequal, as show in
Fig.\ref{obc}(a). An intricate phenomena is that because of the
finite-size effect $\varepsilon_p>\varepsilon_n$ for small $L$,
while under the limit of large $L$, $\varepsilon_p<\varepsilon_n$.
Comparing Fig.\ref{obc} and Fig.\ref{pbc}, our calculation show that
the two $\varepsilon_p$ are equivalent numerically under large $L$
whether for PBC or OBC, so does for $\varepsilon_n$. While for small
$L$, the value of $\varepsilon_p$ is obviously dependent on the
boundary conditions. These findings are consistent with our
assumption that the boundary effects have exponentially decayed
affect with the increment of system size. It means that for VBS
state, geometric entanglement under thermodynamic limit is
independent on the boundary condition. Hence the scaling behavior of
$\varepsilon_p$ with spin $s$ shown in Fig. \ref{obc}(b), is same to
Fig.\ref{pbc}(b) and has the same fitting functions.

\section{Conclusion and Further Discussion}
In conclusion the geometric entanglement is discussed explicitly for
integer spin Valence-Bond-Solid state in this paper. In order to
reduce the optimization in the definition Eq.\eqref{ge}, we point
out that the overlap between the entangled state $\ket{\Psi}$ and
the fully separable state $\ket{\Phi}$ is dependent on the common
symmetry shared by two states. With this important observation, the
fully separable state $\ket{\Phi}$, which maximizes the overlap with
VBS state, can determined explicitly whether for PBC or OBC. For
PBC, $\ket{\Phi}$ is permutational invariant and can be determined
exactly. While for OBC, since the perturbation induced by the
boundary condition, the fully separate state $\ket{\Phi}$ can be
decided when the system size tends to be infinite, and the effect of
boundary condition can be ignored in this case. Furthermore in order
to display the validity of the approximations made to determine
$\ket{\Phi}$ exactly, the numerical simulations have been
implemented, which demonstrates clearly the validity of the
approximations.

An intricate property for the evaluation of GE is that
$\varepsilon_{p(n)}$ is dependent on the parity of the particle
number $L$. The similar feature can also be found for the evaluation
of entanglement entropy in one-dimensional system \cite{lsca06},
however in which case the dependence of the parity of particle
number is because of the boundary conditions. While for VBS state,
the parity dependence of GE is mainly because that the elementary
matrix ${^{p(n)}\bra{\phi}}\otimes g_i$ is traceless. Another
interesting feature for GE is that the $\varepsilon_p$ for PBC and
OBC coincides under large $L$. Compared to the entanglement entropy,
which heavily depends on the boundary conditions\cite{ecp08}, this
feature can be considered as a manifest of the overall character of
multipartite entanglement, and shows that the local boundary
condition would be negligible when one focuses on the global
entanglement in the system.

The important observation for GE is that $\varepsilon_p$ becomes
saturated with the increment of particle number $L$ as shown in
Fig.\ref{pbc}(a) and Fig.\ref{obc}(a). Since $\varepsilon_p$ denotes
the average entanglement per particle, this feature strongly implies
that the total entanglement for the whole system is proportional to
the particle number $L$, and the globe GE $E_G$ is divergent under
$L\rightarrow\infty$. Given the appearance of the finite energy gap
for the system described by the hamiltonian
Eq.\eqref{ham}\cite{arovas88}, this divergency means that there is
hidden symmetry for the hamiltonian Eq.\ref{ham}. Together with the
finding of the hidden $Z_{s+1}\times Z_{s+1}$ topological symmetry
for Eq.\eqref{ham}\cite{kt92,tu08}, it implies that GE could be used
to detect the topological phase transition. Moreover this divergency
also present a clear explanation for the appearance of maximal
localized entanglement in AKLT model\cite{verstraete04}. Since the
degeneracy of ground state of Eq.\eqref{ham} still exists manifested
by the divergence of $E_G$ under $L\rightarrow\infty$, it means that
the long-range correlation could still be founded between any two
particles. Another important observation is that the scaling
behavior of $\varepsilon_p$ with spin number $s$ is obviously
dependent on the parity of spin $s$ shown in Fig.\ref{pbc}(b) and
Fig.\ref{obc}(b). This feature is consistent with the observation
that different SOPs have to be chosen for odd and even spin-$s$ VBS
state in order to identify the different topological
symmetries\cite{ts95}.

Given these points it is believed that GE could include more
information of VBS state than that of entanglement entropy. These
critical differences between GE and entanglement entropy for VBS
state suggest that multipartite entanglement would be more popular
for our understanding of the many-body effects since GE presents an
overall description for many-body systems. Moreover it also hints
that GE could be used to mark the different topological phases, of
which the identification is still a difficult task. Another
intricate points is that the convergence of GE is much slower than
that of entanglement entropy \cite{khh07}. In our own point, this
feature can attribute to the long-distance correlation characterized
by SOP.

Finally, it should point out that the definition and measurement of
multipartite entanglement are a difficult task in
general\cite{bgh10}. By the discussion of GE in VBS state, we should
express our own point that it is efficient  for the construction of
a comprehensive understanding of multipartite entanglement to found
its connection with the diverse physical phenomena, especially in
many-body systems. Hence we wish this work may be useful for this
purpose.

\begin{acknowledgement}
The author(Cui) thanks the helpful discussion with Dr. Chang-shui Yu
and comments from Dr. Hong-Hao Tu. Especially we appreciate the
criticisms from the scholars whose name we doesn't know during the
preparation and submission of this paper. This work is supported by
the Special Foundation of Theoretical Physics of NSF in China, Grant
No. 10747159.
\end{acknowledgement}

\section*{Appendix A}
This point can be verified by noting
\begin{eqnarray}\label{gm}
\text{Tr}[{_p\bra{\Phi}}g_1\cdot g_2\cdots
g_L]=\text{Tr}[\prod_{i=1}^L {^p\bra{\phi}}\otimes g_i]
\end{eqnarray}
With respect of the expressions for $g_i$ (Eq. \eqref{g}) and
${^p_i\bra{\phi}}$ (Eq. \eqref{phi}), the relation above can be
rewritten as a block-diagonal matrix for $s=2k-1(k=1,2,\dots,n)$
\begin{eqnarray}
&&{^p\bra{\phi}}\otimes g_i= c_p\nonumber\\ &&\left(
\begin{array}{cc} \dashbox(160,60){$\begin{array}{cccc}g_{1,1}&
g_{1,3}& \cdots
&g_{1, 2k-1} \\ g_{3,1}& g_{3,3}& \cdots & g_{3,2k-1} \\
\vdots & & &\vdots \\ g_{2k-1,1} & g_{2k-1,3} & \cdots &
g_{2k-1,2k-1}\end{array}$} & 0 \\ 0 &
\dashbox(160,60){$\begin{array}{cccc}g_{2k,2k} & g_{2k,2k-2} &
\cdots
& g_{2k-2,2}\\g_{2k-2,2k}& g_{2k-2,2k-2}& \cdots &g_{2k-2,2} \\
\vdots & & &\vdots \\g_{2,2k}& g_{2,2k-2}& \cdots & g_{2,2}
\end{array}$}\end{array}\right)
\end{eqnarray}
where
\begin{eqnarray*}
g_{p,q}=(-1)^{s+p-1}\sqrt{{_{s}C_{p-1}}{_{s}C_{q-1}}}\sqrt{(s-p+q)!(s+p-q)!}.
\end{eqnarray*}
The first submatrix correspondes to the case both of $p$ and $q$
being odd number, while the second submarix corresponding to even
$p,q$. Noting that ${_{2k-1}C_{2k-2m-1}}={_{2k-1}C_{2m+1 -1}}$, one
conclude that the two submatrices are identical execpt of a negtive
sign. Moreover this matrix is independent of the coordiante $i$ of
particles, and then for odd $L$ the trace of Eq. \eqref{gm} is
always zero.

However although the block-diagonal form still exists for even spin
$s$, the two diagonal blocks do not have the same dimensions. The
conclusion above is not correct for even $s$ VBS state.

The analysis is similar for the evaluation $\epsilon_n$. In this
case ${^n_i\bra{\phi}}\otimes g_i$ can be formulated as a
block-antidiagonal matrix, i.e.
\begin{eqnarray}
&&{^n_i\bra{\phi}}\otimes g_i=c_n\nonumber\\&& \left(
\begin{array}{cc} 0& \dashbox(140,60){$\begin{array}{cccc}g_{1,2}&
g_{1,4}& \cdots
&g_{1, 2i} \\ g_{3,2}& g_{3,4}& \cdots & g_{3,2i} \\
\vdots & & &\vdots \\ g_{2j-1,2} & g_{2j-1,4} & \cdots &
g_{2j-1,2i}\end{array}$} \\
\dashbox(140,60){$\begin{array}{cccc}g_{2,1} & g_{2,3} & \cdots
& g_{2,2j-1}\\g_{4,1}& g_{4,3}& \cdots &g_{4,2j-1} \\
\vdots & & &\vdots \\g_{2i,1}& g_{2i,3}& \cdots & g_{2i,2j-1}
\end{array}$}& 0\end{array}\right)
\end{eqnarray}
where $i=[(s+1)/2], j=[s/2]+1$ (the function $[f]$ denotes the
integer part not more than $f$ ). Then for odd $L$
$\text{Tr}[\prod_{i=1}^L {^n_i\bra{\phi}}\otimes g_i]$ always is
zero.


\begin{thebibliography}{99}
\bibitem{afov07}L. Amico,  R. Fazio, A. Osterloh, and V. Vedral,
Rev. Mod. Phys. {\bf 80}, 517 (2008).

\bibitem{osterloh08} A. Osterloh, e-print at
arXiv:0810.1240[quant-ph].

\bibitem{oaff02} A. Osterloh,  L. Amico, G. Falci and R. Fazio, Nature, {\bf 416}, 608(2002)

\bibitem{on02}T. J. Osborne and  M. A. Nielsen, Phys. Rev. A {\bf 66}, 032110(2002).

\bibitem{ecp08} J. Eisert, M. Cramer and M.B. Plenio, Rev. Mod. Phys. 82, 277(2010) and available at
e-print arXiv: 0808.3773.

\bibitem{multi}O. G\"uhne, G. To\'th and H.J. Briegel, New J. Phys. {\bf 7}, 229(2005); T.R. de Oliveira, G. Rigolin, M.C. de Oliveira and E. Miranda, Phys.
Rev. Lett. {\bf 97}, 170401(2006); H.T. Cui, Phys. Rev. A, 77,
052105(2008);  R. Or\'{u}s, S. Dusuel, J. Vidal, Phys.Rev.Lett.{\bf
100}, 130502(2008); Q.-Q. Shi, R. Or\'{u}s, J. Ove Fjaerestad, H.-Q.
Zhou, New J.Phys. {\bf 12}, 025008(2010); R. Or\'{u}s, T.-C. Wei,
arXiv:0910.2488v2 [cond-mat.str-el]; C.-Y. Huang and F.-L. Lin,
Phys. Rev. A {\bf 81}, 032304 (2010); H.T. Cui,
arXiv:1002.2139v3[quant-ph], accepted by PRA.

\bibitem{odv08}R. Or\'{u}s, S. Dusuel, J. Vidal, Phys.
Rev. Lett. {\bf 101}, 025701 (2008).

\bibitem{wei05}T.C. Wei, D. Das, S. Mukhopadyay, S. Vishveshwara, and P.M. Goldbart, Phys. Rev. A {\bf 71}, 060305(2005).

\bibitem{orus08}R. Or\'{u}s, Phys. Rev. A {\bf 78}, 062332 (2008).

\bibitem{tele}S. Bose, Phys. Rev. Lett. {\bf 91}, 207901 (2003).

\bibitem{gi09} S.M. Giampaolo and F. Illuminati, Phys. Rev. A {\bf 80}, 050301(R) (2009)

\bibitem{lsm61} E. Lieb, T. Schultz, D. Mattis, Ann. Phys. {\bf 16}
407(1961); I. Affleck and E.H. Lieb, Lett. Maht. Phys. {\bf 12},
57(1986).

\bibitem{haldane} F.D.M. Haldane, Phys. Lett. A {\bf 93}, 464(1983);
Phys. Rev. Lett. {\bf 50}, 799(1987).

\bibitem{aklt}I. Affleck, T. Kennedy, E.H. Lieb, and H. Tasaki, Phys.
Rev. Lett. {\bf 59}, 799(1987); Commun. Math. Phys. {\bf 115},
477(1988).

\bibitem{arovas88}D.P. Arovas, A. Auerach, and F.D.M. Haldance,
Phys. Rev. Lett. {\bf 60}, 531(1988).

\bibitem{verstraete04}F. Verstraete, M.A. Mart\'{\i}n-Delgado, and J.I. Cirac, Phys. Rev. Lett. {\bf 92}, 087201(2004).

\bibitem{venuti06}L. Campos Venuti, C.D.E. Boschi, and M. Roncaglia, Phys. Rev. Lett.
{\bf 96}, 247206 (2006).

\bibitem{venuti05}L. Campos Venuti and M. Roncaglia, Phys. Rev. Lett.
{\bf 94}, 207207(2005).

\bibitem{mk89}M. den Nijs and K. Rommels, Phys. Rev. B {\bf 40},
4709(1989).

\bibitem{kt92}T. Kennedy and H. Tasaki, Commun. Math. Phys. {\bf
147}, 431(1992).

\bibitem{oshikawa}M. Oshikawa, J. Phys.: Condens. Matter, {\bf 4},
7496(1992).

\bibitem{ts95}K. Totsuka and M. Suzuki, J. Phys. A: Math. Gen. {\bf 27}, 6443(1994);
J. Phys.: Condens. Matter, {\bf 7}, 1639(1995).

\bibitem{tu08} H.H. Tu, G.M. Zhang, and T. Xiang, J. Phys. A: Math. Theor. {\bf 42} (2009) 2852.

\bibitem{vidal}G. Vidal, J. Mod. Opt. {\bf 47}, 355(2000).

\bibitem{fan04}H. Fan, V. Korepin, and V. Roychowdhury, Phys. Rev.
Lett. {\bf 92}, 027901(2004); H. Fan, V. Korepin, and V.
Roychowdhury, C. Halday, and S. Bose, Phys. Rev. B {\bf 76},
014428(2007).

\bibitem{katsura}H. Katsura, T. Hirano, and Y. Hatsugai1,, Phys. Rev. B {\bf
76}, 012401(2007); H. Katsura, T. Hirano, and V.E. Korepin, J. Phys.
A: Math. Gen. {\bf 41}, 135304(2008);Y. Xu, H. Katsura, and V.E.
Korepin, J. Stat. Phys. {\bf 133}, 347-377(2008).

\bibitem{shimony95}A. Shimony, Ann. NY. Acad. Sci., {\bf 755},
675(1995).

\bibitem{chs00}H. A. Carteret, A. Higuchi and A. Sudbery, J. Math. Phys. {\bf 41},
7932-7939(2000).

\bibitem{bl01}H. Barnum and N. Linden, J. Phys. A: Math. Gen. {\bf
34}, 6787(2001).

\bibitem{wg03}T.C. Wei and P.M. Goldbat, Phys. Rev. A {\bf 68},
042307(2003).

\bibitem{cw07}Y. Cao and A.M. Wang, J. Phys. A: math. and Theo. {\bf
40}, 3507(2007).

\bibitem{wei09}Tzu-Chieh Wei and S. Severini, e-print at arXiv:
0905.0012; M. Hayashi, D. Markham, M. Murao, M. Owari and S.
Virmani, J. Math. Phys. 50, 122104 (2009).

\bibitem{hkwg09}R. H\"ubener, M. Kleinmann, T.C. Wei, O. G\"{u}hne,
Phys. Rev. A {\bf 80}, 032324 (2009) or e-print at arXiv: 0905.4822.

\bibitem{he81}F. T. Hioe and J. H. Eberly, Phys. Rev. Lett. {\bf
47}, 838(1981).

\bibitem{fm91} W.D. Freitag and E. M\"uller-Hartmann, Z. Phys.
B-Condensed Matter, {\bf 83}, 381(1991).

\bibitem{lsca06}N. Laflorencie, Erik S. S{\o}ensen, M.S. Chang and I.
Affleck, Phys. Rev. Lett. {\bf 96}, 100603(2006).

\bibitem{khh07} H. Katsura, T. Hirano and Y. Hatsugai, Phys. Rev. B {\bf 76}, 012401
(2007).

\bibitem{bgh10}C.H. Bennett, A.Grudka, M. Horodecki,
P. Horodecki, and R. Horodecki, arXiv:0805.3060v2 [quant-ph](2010).
\end{thebibliography}
\end{document}